\documentstyle[aps,prl,twocolumn,epsf]{revtex}
\begin{document}
\draft
\wideabs{
\title{Superfluid transition in quasi2D Fermi gases} 
\author{D.S. Petrov${}^{1,2}$ M.A. Baranov${}^{2,3}$, and
G.V. Shlyapnikov${}^{1,2,4}$}
\address{${}^1$ FOM Institute for Atomic and Molecular
Physics, Kruislaan 407,
1098 SJ Amsterdam, The Netherlands \\
${}^2$ Russian Research Center Kurchatov Institute, 
Kurchatov Square, 123182 Moscow, Russia \\ 
${}^3$ Institut f\"ur Theoretische Physik, Universit\"at
Hannover, D-30167
Hannover, Germany \\
${}^4$ Laboratoire Kastler Brossel, 24 rue
Lhomond,
F-75231 Paris Cedex 05, France} 
\date{\today}
\maketitle
\begin{abstract}

We show that atomic Fermi gases in quasi2D geometries 
are promising for achieving superfluidity. In the regime of
BCS 
pairing for weak attraction, we calculate the critical 
temperature $T_c$ and analyze possibilities of increasing
the 
ratio of $T_c$ to the Fermi energy. In the opposite limit,
where a strong coupling leads to the formation of weakly
bound 
quasi2D dimers, we find that their Bose-Einstein condensate 
will be stable on a long time scale.

\end{abstract}
}

Recent progress in trapping and cooling of Fermi isotopes of
K \cite{Jin,Massimo} and Li
\cite{Randy,Christophe,Thomas,Kai} 
has shown the ability to go far below the temperature 
of quantum degeneracy and to manipulate independently the
trapping geometry, 
density, temperature and interparticle interaction. 
The Duke experiment \cite{T} presents intriguing results
on the possibility of achieving a superfluid phase
transition 
in the two-component Fermi gas of $^6$Li.


Two-dimensional Fermi gases have striking features not
encountered in 3D. In the superfluid state, thermal 
fluctuations of the phase of the order parameter strongly
modify the phase coherence properties. The
interaction strength depends logarithmically on the
relative energy of the colliding atoms. For degenerate
Fermi gases this energy is of the order of the Fermi energy
$\varepsilon_F$ which is proportional to the 2D density
$n$. Accordingly, the exponential dependence of the BCS
transition temperature on the interaction strength
transforms into a power law dependence on the density:
$T_c\propto n^{1/2}$ \cite{Miyake,RS}. This suggests a
unique possibility to cross the critical point by
adiabatically expanding a degenerate Fermi gas. Since
the ratio $T/\varepsilon_F$ remains unchanged, the 
temperature scales as $n$ and decreases with 
density faster than $T_c$.

Experimentally it is possible to achieve the quasi2D regime
by confining the atoms in one direction so tightly that the
corresponding level spacing exceeds the Fermi energy. Under
this condition the degenerate Fermi gas is kinematically
two-dimensional. Thus far, this regime has been reached for
Cs atoms \cite{Chu,Christ,Hammes} and for Bose-Einstein
condensates of Na \cite{Ketlowd} and Rb \cite{Ing}. 

In the quasi2D regime the mean-field interaction between
particles exhibits a similar logarithmic dependence on the 
particle energy as in the purely 2D case \cite{Petrov}. 
The amplitude of the $s$-wave scattering turns out to be
sensitive to the strength of the tight confinement
\cite{Petrov}. This opens new handles on manipulations
of the interparticle interaction and superfluid pairing. 

In this Rapid Communication we show that atomic Fermi gases
in quasi2D geometries can become strong competitors of 3D gases
in achieving superfluidity. The ability to increase the
interparticle interaction by tuning the trap frequencies
gives an opportunity to realize a transition from
the standard BCS pairing in the case of weak
attraction to the limit of strong interactions and pairing
in coordinate space. In the latter case one eventually gets a 
dilute system of weakly bound quasi2D dimers of fermionic
atoms, which
can undergo Bose-Einstein condensation. For the BCS case,
we calculate the critical temperature 
$T_c$ to second order in perturbation theory and discuss
possibilities of increasing the ratio $T_c/\varepsilon_F$. 
In the other extreme, we find that the interaction between 
the quasi2D dimers is repulsive, and their collisional 
relaxation and decay are strongly suppressed. This allows us 
to conclude that BEC of these composite bosons will be
stable 
on a long time scale.

We consider an ultracold two-component Fermi gas in the
quasi2D regime and confine ourselves to the $s$-wave
interaction 
and superfluid pairing between atoms of different
components. 
We assume that the
characteristic radius of the interaction potential is much
smaller than the
harmonic oscillator length in the tightly confined
direction,
$l_0=(\hbar/m\omega_0)^{1/2}$, where $m$ is the atom mass,
and $\omega_0$ is the
confinement frequency. Then the interaction problem involves
two length
scales: $l_0$ and the 3D scattering length $a$. For $a<0$
and $|a|\ll l_0$,
there is a peculiar quasi2D weakly bound $s$-state of two
particles, with the
binding energy \cite{Petrov}    
\begin{equation}    \label{epsilon0}
\varepsilon_0=0.915(\hbar\omega_0/\pi)\exp{(-\sqrt{2\pi}l_0/|a|)}\ll
\hbar\omega_0.
\end{equation}
In this case the coupling constant for the intercomponent
interaction takes the form
$g=(4\pi\hbar^2/m)\ln^{-1}(\varepsilon_0/\varepsilon)$, 
where the relative collision energy $\varepsilon$ is assumed
to be either much
smaller or much larger than $\varepsilon_0$ (see
\cite{Petrov} and refs. therein).
As in degenerate Fermi gases one has
$\varepsilon\sim\varepsilon_F$, the 
interaction is attractive ($g<0$) if the density is
sufficiently high and
one satisfies the inequality 
\begin{equation}       \label{ineq}
\varepsilon_0/\varepsilon_F\ll 1.
\end{equation}

Thus, the inequality (\ref{ineq}) is the necessary condition
for the BCS pairing. For finding the critical temperature
$T_c$ 
below which the formation of Cooper
pairs becomes  favorable, we go beyond the simple BCS
approach and proceed
along the lines of the theory developed by Gor'kov and
Melik-Barkhudarov for
the 3D case \cite{Gor'kov}. 

\begin{figure}
\hspace{-0.2cm}
\epsfxsize=\hsize
\epsfbox{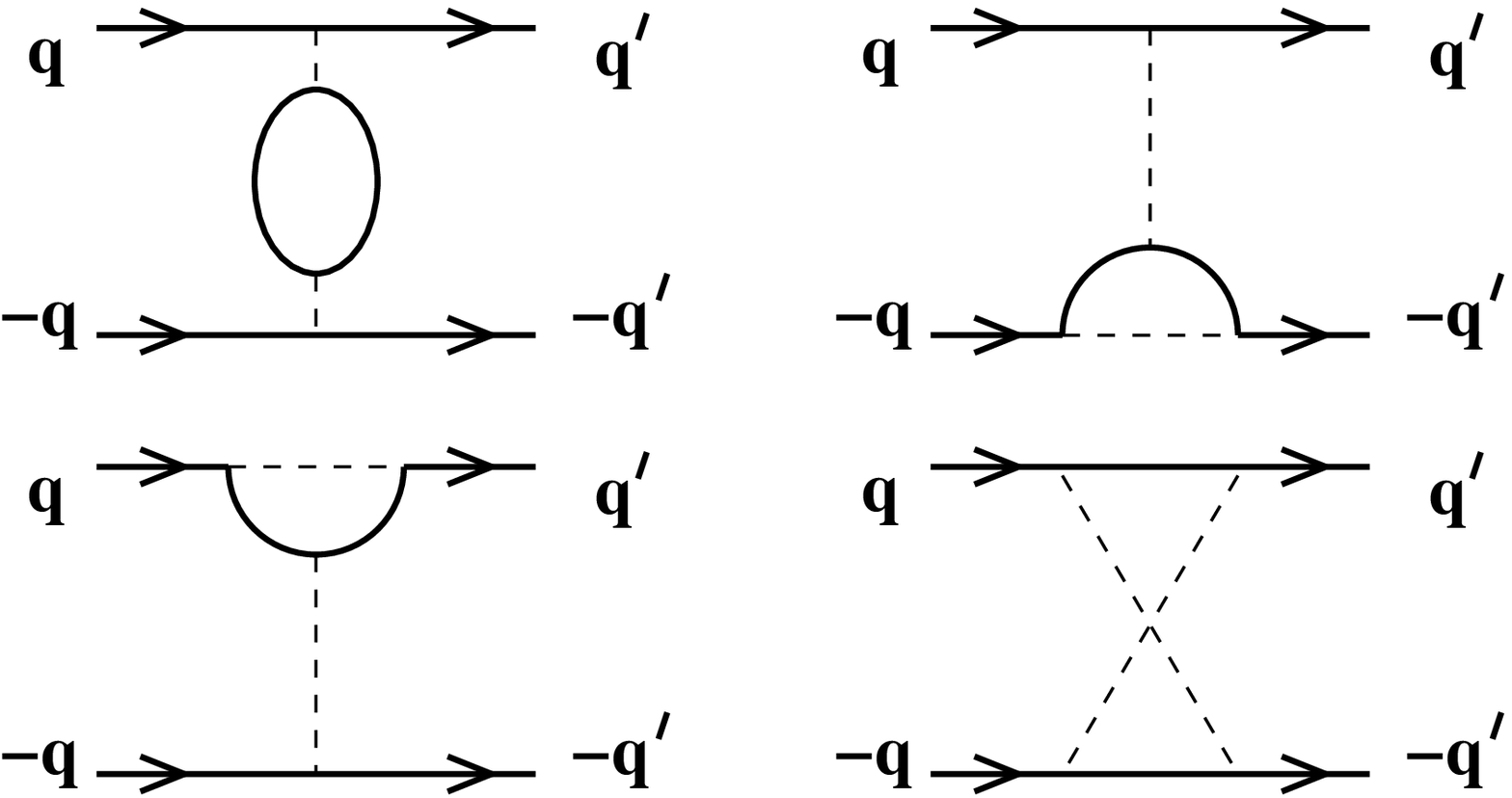}
\caption{\protect
The leading contributions to $\delta V({\bf q},{\bf
q}^{\prime})$.}
\label{Fig1}
\end{figure}

The critical temperature $T_c$ is determined as the highest
temperature for which the linearized equation for the order 
parameter (gap) $\Delta=\langle g\hat\Psi\hat\Psi\rangle$
has a nontrivial solution \cite{Legg}. Assuming that the
quasi2D
gas is uniform in two in-plane directions, the gap equation
in
the momentum space takes the 2D form 
\begin{eqnarray}\label{D} 
\Delta({\bf q})\!  & \approx &
\!-\!\int\Biggl\{\frac{\hbar^2}{m}f(z,{\bf
q,q'}) \left[ K(q^{\prime })+\frac{1}{z-\xi(q^{\prime
})+i0}\right]    \nonumber  \\    
& + \! & \delta V({\bf q},{\bf
q}^{\prime})K(q^{\prime})\Biggr\}\Delta({\bf
q}^{\prime})\frac{{\rm d}^2 q^{\prime}}{(2\pi)^2}, 
\end{eqnarray}     
where $K(q)=(1/2\xi(q))\tanh (\xi(q)/2T)$,
$\xi(q)=\hbar^2q^{2}/2m-\mu$, and
$\mu\approx \varepsilon_F=\pi\hbar^2n/m$ is the chemical
potential. The first term in the rhs of Eq.(\ref{D}) results
from 
the direct interaction between particles, and we
renormalized 
the interaction potential in terms of the scattering
amplitude 
$f$ (vertex function). The latter is a solution of the
quasi2D 
scattering problem. The parameter $z$ has a meaning of the
total 
energy of colliding particles in their center of mass
reference frame.
It is of the order of $\varepsilon_F$ and drops out of the
final 
answer.  The term $\delta V({\bf q},{\bf
q}^{\prime})$ describes
the modification of the interparticle interaction due to the
presence of other
particles (many-body effects). The leading contributions to
this term are second order in the scattering amplitude. They
are
shown in Fig.\ref{Fig1} and correspond to an
indirect interaction between 
two particles when one of them
interacts with a particle-hole pair virtually created from
the ground state
(filled Fermi sea) by the other particle. These second order
contributions are
important for the
absolute value of the critical temperature (preexponential
factor in the 3D
case), whereas higher order terms involving more interaction
events can be 
neglected.     

The amplitude $f$ is independent of the momenta ${\bf q,q'}$
and we will use the 2D relation (see \cite{Popov})  
\begin{equation}       \label{f}
f=4\pi\ln^{-1}\{\varepsilon_0/(-z)\}.
\end{equation}
The last term in the rhs of Eq.(\ref{D}) is a small
correction, since $\delta V\sim f^2$. Accordingly, the
quantity $|f|$ represents a small parameter of the theory. 
In the quasi2D regime ($z\ll \hbar\omega_0$) the motion of
particles 
in the tightly confined direction provides a correction to
Eq.(\ref{f}),
which is $\sim (z/\hbar\omega_0)f^2$ \cite{Petrov}. 
It is much smaller than $\delta V$ and will be  omitted.

Equations (\ref{D}) and (\ref{f}) show that the
momentum dependence of
the order  parameter appears only due to the second order
term that contains
many-body  contributions to the interparticle interaction
$\delta V$. The latter is a function of $p=|{\bf q}+{\bf
q'}|$ 
and rapidly decays for $p>2q_F$, where 
$q_{F}=\sqrt{2m\mu }/\hbar$ is the Fermi momentum. 
For $p\leq 2q_F$ the quantity $\delta V$ is almost constant.
Therefore, one has $\Delta ({\bf
q}')\approx \Delta (q_F)$ in a wide momentum range near the
Fermi surface. 
Then, for $q=q_F$ a direct integration of Eq.(\ref{D})
yields 
\begin{eqnarray}
\Delta (q_{F}) & = & -\frac{f(z)}{4\pi }\Delta(q_F)\ln
\left[\frac{(-2\mu
z)}{\pi^2 T^2}\exp(2\gamma )\right]   \nonumber \\
& - & \delta V(q_{F},q_{F})\frac{m}{2\pi\hbar^2}
\ln  \left( C\frac{\mu }{T}\right)\Delta (q_{F}),  \nonumber 
\end{eqnarray}
where $\gamma\approx 0.5772$ is the Euler constant, and $C$
is 
a numerical factor determined by the momentum 
dependence of $\Delta$ and $\delta V$. The calculation of
$\delta 
V(q_{F},q_{F})$ is straightforward and gives $\delta 
V(q_{F},q_{F})=(\hbar^2/2\pi m )f^2(2\mu )$. Then, using
Eq.(\ref{f}) we
obtain the critical temperature  
\[ 
T_{c}\approx (2\mu/\pi)\exp\left(\gamma -1-|2\pi
{\rm
Re}f^{-1}(2\mu)|\right).  \] 
The exponent in this equation should be large and the
quantity ${\rm
Re}f^{-1}(2\mu)$ should be negative. As the chemical
potential is $\mu\approx
\varepsilon_F$, from Eq.(\ref{f}) one sees that these
requirements are reached
under the condition (\ref{ineq}). Using
Eqs.~(\ref{epsilon0})
and (\ref{f}) the critical temperature takes the form   
\begin{equation}      \label{T_c} 
T_c\!=\!\frac{\gamma\sqrt{2\varepsilon_0\varepsilon_F}}{\pi
e}\!=
0.16\sqrt{\varepsilon_F\hbar\omega_0}\exp\left(\!\!
-\sqrt{\frac{\pi}{2}}\frac{l_0}{|a|}\!\right)
\ll\varepsilon_F.
\end{equation}  
The relative correction to this result is of the order of
$1/|\ln(\varepsilon_0/\varepsilon_F)|\ll 1$.

Note that Eq. (\ref{T_c}) predicts by a factor of $e$
smaller value for the critical temperature than a simple BCS 
calculation \cite{BCS}. This
means that 
the attractive interaction between particles becomes weaker
once we
take into account the polarization of the medium.

The ratio $T_c/\varepsilon_F$ is not necessarily very small.
For example, using Feshbach resonances the scattering length
is tunable over a wide interval of negative values 
\cite{Jin,Thomas,Kai}.
Keeping the exponential term equal to $0.05$ in
Eq.(\ref{T_c}), with 
$\omega_0\sim 100$ kHz we obtain $T_c/\varepsilon_F\sim
0.1$ for 2D densities $n\sim 10^9$ cm$^{-2}$ ($T_c\sim 40$
nK). 

As in the purely 2D case \cite{Miyake,RS}, the transition
temperature 
$T_c\propto n^{1/2}$ and the ratio $T_c/\varepsilon_F$
increases 
with decreasing density as $n^{-1/2}$.
This is a striking difference from the 3D case, where this
ratio decreases exponentially with density. 
In the presence of the in-plane confinement, one can
approach
the BCS transition in a degenerate Fermi gas by
adiabatically
expanding the quasi2D trap in the in-plane direction(s). As
the degeneracy parameter $T/\varepsilon_F$ is conserved in
the 
course of the adiabatic expansion, the ratio $T/T_c$ will 
decrease as $n^{1/2}$. Equations (\ref{epsilon0}) and
(\ref{T_c}) 
also show that one can increase $\varepsilon_0$ and 
$T_c/\varepsilon_0$ by tuning $|a|$ to larger
values or by making the tight confinement stronger and thus
decreasing $l_0$. 

What happens if $\varepsilon_0$ and $\varepsilon_F$ become
comparable 
with each other, i.e. one reaches the quasi2D resonance for
two-body collisions? Then Eq.(\ref{T_c}) leads to
$T_c\sim\varepsilon_F$ and is no
longer valid. In fact, for $\varepsilon_0>\varepsilon_F$
the formation of
bound quasi2D dimers of distinguishable fermions becomes
energetically
favorable and one encounters the problem of Bose-Einstein
condensation of
these bosonic molecules. Thus, an increase of the ratio
$\varepsilon_0/\varepsilon_F$ from small to large values is
expected to
provide a transformation of the BCS pairing to molecular
BEC. This type of crossover has been discussed in literature 
in the context of superconductivity
\cite{Eag,Leg,Noz,Rand,RS} and in relation to superfluidity
in 2D films of $^3$He \cite{Miyake,MYu}. 
The idea of using a Feshbach resonance for achieving a 
superfluid transition in the BCS-BEC crossover regime 
in ultracold 3D Fermi gases has been
proposed in refs. \cite{Holland1,Tim}.

We will not consider the crossover regime and
confine ourselves to the
limiting case of molecular BEC
($\varepsilon_0\gg\varepsilon_F$). A
subtle question is related to the stability of the expected
condensate, which depends on the interaction between the
molecules. For the
repulsive interaction one will have a stable molecular BEC,
and the attractive
interaction should cause a collapse.

The molecule-molecule scattering is a
4-body problem described by the Schr\"odinger equation
\begin{eqnarray}\label{4bodySchr}
&&\Biggl[-\frac{\hbar^2}{m}\left(\nabla_{{\bf
r}_1}^2+\nabla_{{\bf
r}_2}^2\right)-\frac{\hbar^2}{2m}\nabla_{\bf
R}^2+U(r_1)+U(r_2)\nonumber\\
&&+\sum_\pm U\left(\frac{{\bf r}_1+{\bf r}_2}{2}\pm{\bf
R}\right)-E\Biggr]\Psi({\bf r}_1,{\bf r}_2,{\bf R})=0.
\end{eqnarray}
Here ${\bf r}_1$ is the distance between two given
distinguishable
fermions, ${\bf r}_2$ is the distance between the other
two, ${\bf R}$ is the distance between the centers of mass
of
these pairs, and $U$ is the interatomic potential. The total
energy is 
$E=-2\varepsilon_0+\varepsilon$, with $\varepsilon$ being
the relative
molecule-molecule kinetic energy.  

The interaction between molecules is present only at 
intermolecular distances of the order of or smaller than
the size of a molecule $d_{*}=\hbar/\sqrt{m\varepsilon_0}$. 
Therefore, at energies $\varepsilon\ll\varepsilon_0$ the 
scattering between molecules is dominated by the $s$-wave
channel and can be analyzed on the basis of the solution of
Eq.(\ref{4bodySchr}) for
$\varepsilon=0$. For large $R$ the
corresponding wavefunction is $\Psi({\bf r}_1,{\bf r}_2,{\bf
R})\approx K_0(r_1/d_{*})K_0(r_2/d_{*})\ln(\alpha
R/d_{*})$, where the decaying Bessel function
$K_0(r_{1,2}/d_{*})$
represents the 2-body bound state. The parameter $\alpha$ is
a 
universal constant which can be found by matching the
quantity
$\ln(\alpha R/d_{*})$ with the solution of
Eq.(\ref{4bodySchr}) 
at short distances. Finally, matching $\ln(\alpha R/d_{*})$
with the
wavefunction of free
relative motion of two molecules at distances $d_{*}\ll R\ll
\Lambda_{\varepsilon}$, where
$\Lambda_{\varepsilon}=\hbar/\sqrt{m\varepsilon}$ is their
de Broglie wavelength, we obtain the coupling constant
(scattering
amplitude) for the interaction between  molecules:
\begin{equation}      \label{gmol}
g_m=(2\pi\hbar^2/m)\ln^{-1}(2\alpha^2
e^{-2\gamma}\varepsilon_0/\varepsilon)>0;\,\,\,\,\,\,\,\varepsilon\ll\varepsilon_0.
\end{equation}
 
A precise value of $\alpha$ is not important
as it gives rise to higher order corrections in
Eq.(\ref{gmol}). 
However, in order to make sure that this constant is
neither anomalously large nor anomalously small we have
integrated 
Eq.(\ref{4bodySchr}) numerically. For this purpose, it is
convenient to transform Eq.(\ref{4bodySchr}) into an
integral equation 
for a function which depends only on three 
independent coordinates. This has been done by using the
method
of ref. \cite{Petrov3body}. Our calculations lead to
$\alpha\approx 1.6$. 
They show the absence of 4-body weakly bound
states and confirm an intuitive picture that 
the interaction between two
molecules can 
be qualitatively represented by means of a purely repulsive
potential 
with the range $\sim d_{*}$. For the interaction between
Bose-condensed
dimers, in Eq.(\ref{gmol}) one has 
$\varepsilon=2n_mg_m\ll\varepsilon_0$, where $n_m$ is the
density of the dimers (see \cite{Petrov} and refs. therein). 
We thus conclude that a Bose condensate of these weakly
bound
dimers is stable with respect to collapse. 

The 2D gas of bosons becomes Bose-condensed below the 
Kosterlitz-Thouless transition temperature $T_{KT}$
\cite{KT}
which depends on the interaction between particles. 
According to the recent quantum Monte Carlo simulations
\cite{Prokof'ev}, for the 2D gas with the coupling constant
(\ref{gmol}) the Kosterlitz-Thouless temperature is given by
\begin{equation}\label{KTTtemp}
T_{KT}=(\pi\hbar^2
n_{m}/m)\ln^{-1}\left[(\eta/4\pi)\ln\left(1/n_{m}d_{*}^2\right)\right],
\end{equation} 
where the numerical factor $\eta\approx 380$. For
$\varepsilon_F\ll\varepsilon_0$, the density of dimers
$n_m\approx n/2$ and
the parameter  $(1/n_md_*^2)\approx
2\pi\varepsilon_0/\varepsilon_F$. Then,
for $\varepsilon_0/\varepsilon_F=10$, Eq.(\ref{KTTtemp})
gives $T_{KT}/\varepsilon_F\approx 0.1$ and at densities
$10^8$ cm$^{-2}$ the transition temperature in the case of
$^6$Li is
$T_{KT}\approx 30$ nK.  

The weakly bound dimers that we are considering are
molecules in the highest
rovibrational state and they can undergo collisional
relaxation and decay. 
The relaxation process occurs in pair dimer-dimer or
dimer-atom collisions. It produces diatomic molecules in
deep bound
states and is
accompanied by a release of the kinetic energy. The size of
these deeply bound
molecules is of the order of the characteristic radius of
the interatomic potential
$R_e\ll l_0$, and their internal properties are not
influenced by the tight
confinement.
Therefore, the
relaxation can be treated as a 3D process and it requires
the presence of at
least three fermionic  atoms at distances $\sim R_e$ between
them. Since at least two
of them are identical, the relaxation probability acquires a
small factor $(kR_e)^2$  compared to the case of bosons, 
where $k\sim 1/d_*=\sqrt{\varepsilon_0/\hbar\omega_0}/l_0$
is a characteristic momentum of atoms. The 3D density of
atoms in the quasi2D
geometry is $\sim n/l_0$. Thus, qualitatively, the inverse
relaxation time can
be written as $\tau_{\rm rel}^{-1}\sim \alpha_{\rm rel}n
(R_e/l_0)^2(\varepsilon_0/\hbar\omega_0)/l_0$,
where $\alpha_{\rm rel}$ is the relaxation rate constant for
the highest
rovibrational states of 3D molecules of two bosonic atoms.
We estimate $\tau_{\rm rel}$ keeping in mind the recent
measurements for Rb$_2$ molecules \cite{Heinzen} which give 
$\alpha_{\rm rel}\approx 3\times 10^{-11}$ cm$^3$/s. 
For $l_0$ in the interval from $10^{-5}$ to $10^{-4}$ cm,
the suppression factor 
$(R_e/l_0)^2(\varepsilon_0/\hbar\omega_0)$ 
ranges from $10^{-3}$ to $10^{-5}$ and at
2D densities $n\sim 10^8$ cm$^{-2}$ we find the relaxation
time $\tau_{\rm
rel}$ of the order of a second or larger.  

Dimer-dimer pair collisions can lead to the formation
of bound trimers, accompanied by a release of one of the
atoms. The formation of deeply bound trimer states will be
suppressed at least in the same way as the relaxation
process discussed above. Therefore, it is important that 
there are no weakly bound trimers in (quasi)2D. Just as
in 3D
\cite{Efimov}, this can be
established by using the zero-range approach
($R_e\rightarrow 0$). We have
performed this analysis along the lines of the 3D work
\cite{Petrov3body}.
Qualitatively, the symmetry of the 3-fermion system
containing two identical
fermions provides a strong centrifugal repulsion that
does not allow the presence of
3-body bound states. This is in contrast to 2D bosons where
one has two fully symmetric trimer bound states
\cite{Nielsen}. 

Thus, the life-time of quasi2D
dimers of fermionic
atoms is rather long and one easily estimates that it
greatly exceeds the
characteristic time of elastic collisions. One can even
think of achieving BEC in the initially non-condensed gas of
dimers
produced out of a
non-superfluid atomic Fermi gas under a decrease of $n$ or
$l_0$.  

In conclusion, we have found the temperature of superfluid
phase transition 
in two-component quasi2D Fermi gases. Our results are
promising for achieving
this transition in both the regime of BCS pairing and the
regime of BEC of
weakly bound dimers.

We acknowledge fruitful discussions with C. Salomon, A.
Mosk, C. Lobo, and J.T.M. Walraven. This work was
financially 
supported by the the Dutch Foundations NWO and FOM, by
Deutsche
Forschungsgemeinschaft (DFG), by INTAS, and by the Russian
Foundation for Basic Research. Le LKB est UMR 8552 du CNRS, 
de l'ENS et de l'Universit\'{e} P. et M.~Curie.

\end{document}